\documentclass[final,3p,times,twocolumn]{elsarticle}

\usepackage{amssymb}

\usepackage[utf8]{inputenc} 
\usepackage[T1]{fontenc}    
\usepackage{url}            
\usepackage{booktabs}       
\usepackage{amsfonts}       
\usepackage{nicefrac}       
\usepackage{microtype}      
\usepackage{lipsum}		
\usepackage{graphicx}
\usepackage{siunitx}
\usepackage{amsmath}
\usepackage{float}
\usepackage[hidelinks]{hyperref}

\journal{Optics Communications}

\begin{document}

\begin{frontmatter}



\title{Deep neural networks for efficient phase demodulation in wavelength shifting interferometry}

\author[a]{Jacob Black}
\author[a]{Shichao Chen}
\author[a]{Joseph G. Thomas}
\author[a]{Yizheng Zhu \corref{cor1}} 
\ead{yizhu1@vt.edu}
\cortext[cor1]{Corresponding author}

\address[a]{Bradley Department of Electrical and Computer Engineering
Virginia Tech
Blacksburg, VA 24060 }

\begin{abstract}
Analytical phase demodulation algorithms in optical interferometry typically fail to reach the theoretical sensitivity limit set by the Cram\'er-Rao bound (CRB). We show that deep neural networks (DNNs) can perform efficient phase demodulation by achieving or exceeding  the CRB by significant margins when trained with new information that is not utilized by conventional algorithms, such as noise statistics and parameter constraints. As an example, we developed and applied DNNs to wavelength shifting interferometry. When trained with noise statistics, the DNNs outperform the conventional algorithm in terms of phase sensitivity and achieve the traditional three parameter CRB. Further, by incorporating parameter constraints into the training sets, they can exceed the traditional CRB. For well confined parameters, the phase sensitivity of the DNNs can even approach a fundamental limit we refer to as the single parameter CRB. Such sensitivity improvement can translate into significant increase in single-to-noise ratio without hardware modification, or be used to relax hardware requirements. 

\end{abstract}


\end{frontmatter}


\section{Introduction}
Microscopy techniques that measure the phase distribution of a specimen to generate high contrast images are generally referred to as quantitative phase imaging (QPI) \cite{pope1}. Some examples include holographic phase microscopy \cite{Cuche:99}, spatial light interference microscopy \cite{Wang:11}, optical coherence phase microscopy \cite{GILLIES2018126}, and optical diffraction tomography \cite{KIM2018160}. These and other techniques are reviewed in \cite{NADEAU20181}. Recently, advances have been made in applying deep neural networks (DNNs) to QPI systems \cite{qpiML, phaseStain, physicsDNN}. In particular, DNNs offer an effective solution for holographic image reconstruction \cite{prHolo}, solving inverse problems in lensless imaging \cite{Sinha:17}, and phase recovery from simulated intensity data \cite{Kemp_2018}. Outside of QPI, DNNs have found success in an increasingly diverse field of imaging techniques such as sparse-angle x-ray reconstruction \cite{xRayTomog}, segmentation in magnetic resonance imaging \cite{rpNET}, and super-resolution imaging \cite{superRes1}. These results suggest that DNNs are effective in diverse data processing. 

In this work we focus on the temporal phase sensitivity properties of DNNs in the context of QPI. This sensitivity can be used to evaluate and compare different QPI systems, and is bounded from below by the Cram\'er-Rao bound (CRB) \cite{Chen:17, cramerOrig} . The CRB, a fundamental result in mathematical statistics, provides a basis for evaluating the efficiency of a signal processing algorithm (estimator) \cite{ chengshuaiJSTQE, leongarcia08}.
Reaching the CRB has both theoretical and practical benefits. Increased sensitivity allows for smaller changes in OPL to be detected accurately. Significantly, any sensitivity improvement represents that same improvement squared in signal-to-noise ratio (SNR) \cite{chengshuaiMLE}. For example, a factor of 3 increase in sensitivity results in a factor of 9 increase in SNR, nearly an order of magnitude. By making use of the efficient DNNs developed in this work, even greater increases are possible for experimental conditions where the analytical approach struggles. 

Current analytical algorithms used to calculate OPL in wavelength shifting interferometry (WSI), and indeed most other QPI techniques, do not reach the sensitivities predicted by the CRB \cite{Chen:17}. Using WSI as an example, we demonstrate through simulation and experiment that by incorporating the shot-noise limited nature of our system during the training process, DNNs are capable of achieving the traditional  three parameter CRB developed in \cite{Chen:17}. We also discuss an even tighter lower bound, the single parameter CRB (SPCRB), for the case where the OPL is the only unknown in the signal model (see Eq. \eqref{eq:int1}). We then demonstrate that DNNs can be trained to closely approach the SPCRB by further taking advantage of the parametric constraints on our signal model. Therefore, we show that not only can DNNs exceed the CRB, but they can also approach the SPCRB.

The structure of this work is as follows. A brief introduction to WSI and the sensitivity evaluation framework is given in Section \ref{sec:bGround}. Our particular DNNs, the training process, and the simulation results are described in Section  \ref{sec:DNN}. The experimental results are presented in Section \ref{sec:exp}. We summarize and discuss the implications of our findings in Section \ref{sec:conclusions}.

\begin{figure}
    \centering
    \includegraphics[width=\linewidth]{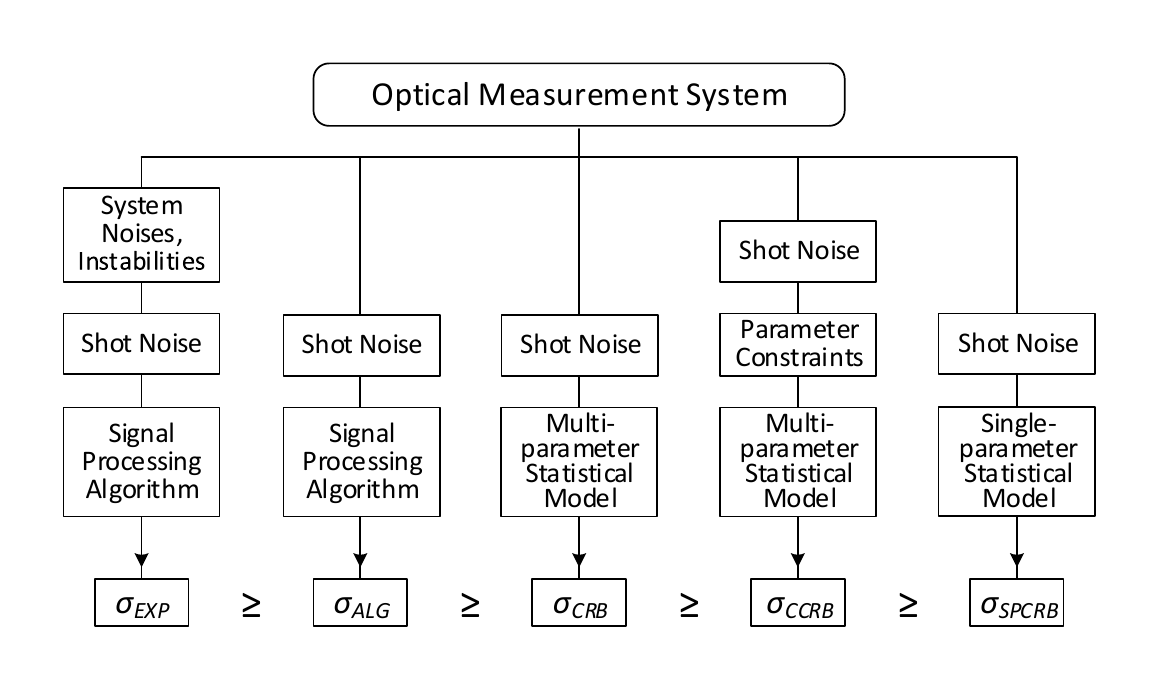}
    \caption{Graphical representation of the general hierarchy of sensitivities for QPI systems. Each block represents a corruption or condition which alters the resulting sensitivity.}
    \label{fig:sensEval}
\end{figure}
\section{Theory of OPL sensitivity evaluation for interferometry}
\label{sec:bGround}
\subsection{Wavelength Shifting Interferometry}
We will focus on demonstrating the performance of our DNNs for OPL demodulation in the context of 4-band WSI \cite{Chen2:16}. We briefly review the theory necessary for a complete sensitivity analysis. In 4-band WSI, four noisy interferograms are captured by the camera, labeled $I_n$. These $I_n$ correspond to the four different wavelength bands produced by a swept laser source. The statistical properties of the $I_n$ are experimentally verified to be shot noise-limited \cite{Chen:17}. In fact, many modern QPI systems satisfy this condition thanks to improvement in system design and the noise performance of cameras. Then, the mean (noise-free) intensity $\Bar{I}_n$, can be expressed in terms of the OPL difference at a given pixel, $L(x,y)$, between the two arms of the interferometer as \cite{goodman_2015}:
\begin{equation}
\label{eq:int1}
    \Bar{I}_n(L(x,y)) = \alpha[1 +  V \cos(k_n L(x,y))], \quad n = 1, 2, 3, 4
\end{equation}
where $\alpha$ is a DC term in analog-to-digital units (ADUs), $V$ is the visibility, and $L(x,y)= L_0 + L_S(x,y)$ with $L_0$ being the OPL difference between the two arms of the interferometer, and $L_S(x,y)$ being the OPL distribution of the sample \cite{Chen:16}. In our WSI, the $k_n$ are evenly spaced with a spacing of $\Delta k$. This gives rise to a constant phase shift, $\Delta \phi = \Delta k L(x,y)$ between adjacent wavenumbers. This constant phase shift, although unknown due to the unknown $L(x,y)$, allows for the application of the well-known Carr\'e equation, which gives the relative OPL in terms of $I_n$ \cite{carrePaper, Chen2:16}:

\begin{multline}
    \label{eq:carre}
    L' = \\ 
    \frac{1}{k_0} \tan^{-1} \left \{\frac{\sqrt{[3(I_2 - I_3) - (I_1 - I_4)](I_2 - I_3 + I_1 - I_4)}}{(I_2 + I_3 - I_1 - I_4) \times \operatorname{sgn}(I_2 -I_3)} \right \}
\end{multline}
where the $'$ refers only to the OPL obtained via  Eq. \eqref{eq:carre}, which is a wrapped version of $L(x,y)$. We will use  Eq. \eqref{eq:carre} in combination with the CRB to evaluate the performance of our DNNs. 

\subsection{Sensitivity Evaluation for QPI Systems}
\subsubsection{Experimental Sensitivity, Algorithm Sensitivity and the CRB}
There is a 5-tier hierarchy of sensitivities as illustrated in Fig. \ref{fig:sensEval}, which is expanded from an earlier 3-tier framework for quantitatively evaluating the  efficiency of a QPI system \cite{Chen:17}. Each sensitivity is the result of a different set of factors that affect system performance. We start in middle with the usual, multi-parameter Cram\'er-Rao bound, denoted by $\sigma_{CRB}$. It is the fundamental sensitivity limit when each of the several variables $\alpha$, $V$, and $L$ in Eq. \eqref{eq:int1} are completely unknown.  $\sigma_{CRB}$ represents the minimum possible value of sensitivity for any unbiased estimator of the parameters of interest, which is OPL in this case. The CRB depends only on the fundamental physical process, i.e. Eq. \eqref{eq:int1}, and the noise distribution that the $I_n$ is corrupted by. For the shot noise-limited case, the noise is Poisson distributed \cite{yariv_yeh_2009}. We obtain $\sigma_{CRB}$ by directly calculating the Fisher information matrix, $J$, and taking particular components of its inverse \cite{leongarcia08}. A calculation of $J$ for 4-band WSI is given by \cite{Chen:17}

\begin{equation}
    \label{eq:fim}
    J =\left(
    \begin{smallmatrix}
    \sum^4_{n=1} \frac{g}{\Bar{I}_n} & \sum^4_{n=1} \frac{g \cos (k_n L)}{\Bar{I}_n} & \sum^4_{n=1} \frac{-g k_n \alpha V \sin (k_n L)}{\Bar{I}_n} \\
    \sum^4_{n=1} \frac{g \cos (k_n L)}{\Bar{I}_n} & \sum^4_{n=1} \frac{g \cos^2 (k_n L)}{\Bar{I}_n} & \sum^4_{n=1} \frac{-g k_n \alpha V \sin (2 k_n L)}{2\Bar{I}_n} \\
    \sum^4_{n=1} \frac{-g k_n \alpha V \sin (k_n L)}{\Bar{I}_n} & \sum^4_{n=1} \frac{-g k_n \alpha V \sin (2k_n L)}{2\Bar{I}_n} & \sum^4_{n=1} \frac{g k^2_n \alpha^2 V^2 \sin^2 (k_n L)}{\Bar{I}_n}
    \end{smallmatrix}
    \right),
\end{equation}

\begin{flushleft}
    where $g$ is the conversion gain associated with the camera in number of electrons per ADU. By definition, the CRB follows as
    
\end{flushleft}

\begin{equation}
    \label{eq:crbDef}
    \sigma_{CRB} = \sqrt{{(J^{-1})_{33}}}
\end{equation}

In practical signal processing, we must also consider the sensitivity associated with a given demodulation algorithm, such as Eq. \eqref{eq:carre}. This sensitivity is referred to as the algorithm sensitivity (ALG) and is defined as $\sigma_{ALG} = \sqrt{\operatorname{Var}(L')}$.  In general, a particular demodulation algorithm degrades the sensitivity when compared to the CRB, as suggested in Fig. \ref{fig:sensEval}. This is due to the fact that  Eq. \eqref{eq:carre} represents an unbiased estimator of $L$. Hence the estimate, $L'$, must come with some variance greater than or equal to that given by the CRB, Eq. \eqref{eq:crbDef}. The equality typically does not hold for algorithms such as Eq. \eqref{eq:carre} because it does not consider noise statistics. We can approximate $\sigma_{ALG}$ by expanding Eq. \eqref{eq:carre} in a Taylor series and directly calculating the variance as detailed in \cite{Chen:17}.      

Finally, there is the experimental sensitivity, $\sigma_{EXP}$, which is the sensitivity measured in practice when using a particular algorithm to obtain $L'$ from raw intensity data. $\sigma_{EXP}$ can be obtained by recording many interferograms of the same sample (e.g. a blank sample), and taking the standard deviation of the experimental OPL, i.e $\sigma_{EXP}$. This sensitivity is degraded further from $\sigma_{ALG}$ by all possible environmental noises, system noises, and instabilities present in a real QPI system. Therefore, in general, we have:
\begin{equation*}
    \sigma_{EXP} \geq \sigma_{ALG} \geq \sigma_{CRB}.
\end{equation*}

\subsubsection{Constrained and Single Parameter CRB}
In the context of this study, we introduce two more levels of sensitivities to the original framework. Again, it is important to emphasize that  Eq. \eqref{eq:crbDef} is the CRB for the case of estimating $L$ without any knowledge of $\alpha$ and $V$. In practice, it is often possible to predict or constrain the values of the ancillary parameters $\alpha$ and $V$ for a given sample and experimental setup. If this partial knowledge is incorporated, a tighter bound known as the constrained CRB (CCRB) becomes the limit for the sensitivity \cite{cCRB}. This informed CCRB is necessarily better than or equal to the uninformed CRB. It is, however, extremely challenging to represent the CCRB analytically, as deriving analytical expressions that account for the partial knowledge of $\alpha$ or $V$ is not at all straightforward as compared to the CRB and its traditional Fisher information approach. Nonetheless, we will show later that DNNs can be used to achieve CCRB.

Further, in the extreme case where we have complete knowledge of $\alpha$ and $V$, our model consisting of the three unknowns in Eq. \eqref{eq:int1} is reduced to a single unknown, $L$. We will refer to this new lower bound as the single parameter CRB (SPCRB).

In general, the SPCRB will be much lower than the bound given by Eq. \eqref{eq:crbDef} \cite{cCRB}. We can calculate $\sigma_{SPCRB}$ directly from  Eq. \eqref{eq:fim} by eliminating rows and columns of $J$ relating to $\alpha$ or $V$ since they are known. This is equivalent to repeating the original derivation for $J$ assuming $L$ is the only unknown. This reduces $J$ to a single component, $J_{33}$, and $\sigma_{SPCRB}$ is thus

\begin{equation}
    \label{eq:spcrbDef}
    \sigma_{SPCRB} = \sqrt{1/J_{33}} = \frac{1}{\sqrt{\sum^4_{n=1} \frac{g k^2_n \alpha^2 V^2 \sin^2 (k_n L)}{\Bar{I}_n}}}.
\end{equation}

This leads to the complete inequality in Fig. \ref{fig:sensEval}:
\begin{equation*}
    \sigma_{EXP} \geq \sigma_{ALG} \geq \sigma_{CRB} \geq \sigma_{CCRB} \geq  \sigma_{SPCRB}.
\end{equation*}

\begin{figure}[!ht]
    \centering
    \includegraphics[width=\linewidth]{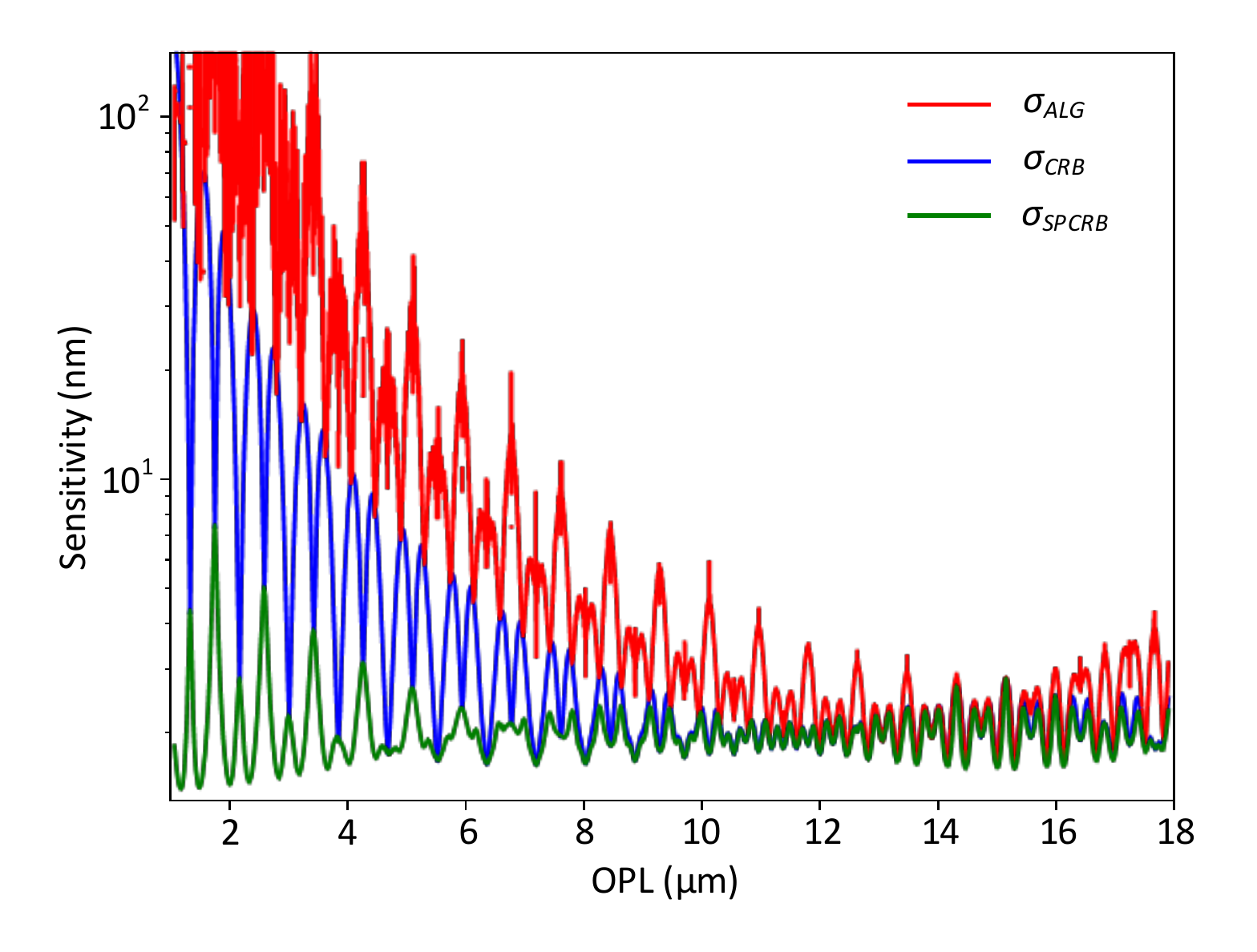}
    \caption{Theoretical sensitivity curves for the ALG, CRB, and the SPCRB. }
    \label{fig:crb_alg_spcrb}
\end{figure}

As an example, $\sigma_{SPCRB}$, $\sigma_{CRB}$, and $\sigma_{ALG}$ are plotted together in Fig. \ref{fig:crb_alg_spcrb} with $\alpha$ and $V$ being $128$ ADU and $0.7$, respectively. $L$ is then swept across $[1, 18]$ $\SI{} {\micro\meter}$. It is clear that the Carr\'e equation (ALG) only approaches the limits predicted by the Cram\'er-Rao bound(s) between approximately 14 and 16 $\SI{}{\micro \meter}$. Outside of this relatively narrow region, $\sigma_{ALG}$ degrades significantly, in some places more than an order of magnitude, when compared with the CRB. This suggests that there is much room for improvement, particularly towards shorter OPLs, where the difference between $\sigma_{ALG}$ and the theoretical minimum become increasingly large. Also note that $\sigma_{CRB} > \sigma_{SPCRB}$ in a similar fashion, as expected.

\section{Deep Neural Network Analysis}
\label{sec:DNN}
\subsection{Architecture}

\begin{figure}[!ht]
    \centering
    \includegraphics[width=\linewidth]{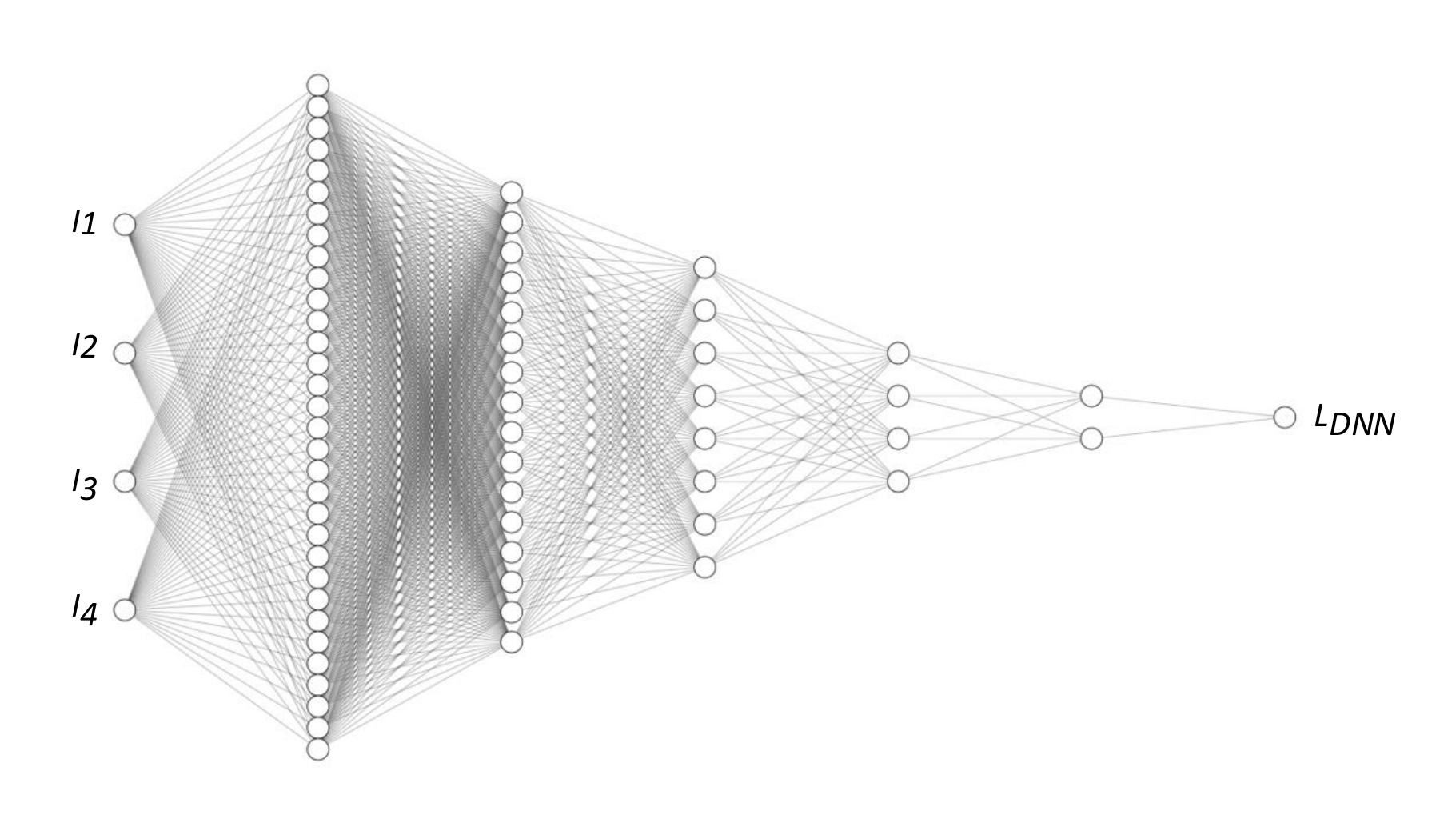}
    \caption{Representative fully connected DNN diagram, excluding dropout layers. Each layer's nodes have a direct connection to each node in the next layer. The output is the total OPL ($L_o + L_S$), labeled $L_{DNN}$.}
    \label{fig:network}
\end{figure}

Deep neural networks are a class of algorithms capable of effectively modeling highly nonlinear processes. This is accomplished by providing training data as input to the network, and the output of the network is then compared against expected outcomes (the truth data). This training strategy is referred to as supervised learning as there is an expected output for a given input. In our case, the input to the DNN will be the interferograms $I_n(x,y)$ from each pixel on the camera, and the output will be the OPL. After training, the output of our DNN, $L_{DNN}$, will be shown to accurately obtain $L$ in Eq. \eqref{eq:int1} with improved sensitivity relative to traditional techniques (see Figs. \ref{fig:simResults1} and \ref{fig:histo1}). 

For this application, we chose to use a traditional, fully connected DNN consisting of 9 layers. Figure \ref{fig:network} shows a representative plot of our network's structure (dropout layers excluded) with the weights, $\Theta$, denoted by the lines connecting all of the nodes together. Each layer contains half of the previous layer's number of neurons (nodes connecting all of the lines). Additionally, there are two dropout layers included immediately after the first hidden layer and prior to the output layer with rate $0.01$ to help prevent over fitting. The activation function for each layer is the standard $\operatorname{sigmoid}$ function $S(x) = 1/(1+e^{-x})$, while the loss function is the usual mean-squared error defined by
\begin{equation*}
    E(\Theta) = \frac{1}{N} \sum_{i=1}^{N} || L^{(i)}(\Theta)_{DNN} - T_Y^{(i)}||^2,
\end{equation*}
where $L^{(i)}(\Theta)_{DNN}$ is the neural network output for the $i^{th}$ training input, and $T_Y^{(i)}$ is the corresponding ground truth value. $E(\Theta)$ represents the objective function being minimized during the training process, and it is emphasized that it is only a function of the weights in the network. Our networks were implemented in Python using the Keras \cite{chollet2015keras} deep learning library with Tensorflow GPU, and the ADAM optimizer was used \cite{Kingma2015AdamAM} with learning rate $\gamma = 0.0005$. 

\subsection{Training Data Generation}
\label{subSec:TDG}
Consider a particular pixel of an $M\times N$  pixel camera located at $(m, n)$, where $ 0 \leq m \leq M$ and $0 \leq n \leq N$. For the purpose of calculating $L(m, n)$ from the four noise-corrupted intensities $\mathbf{I}(m, n) = [I_1, I_2, I_3, I_4]$, we require our training data to accurately model the data from our WSI system. To this end, we assume each $I_n$ takes on integer values in the range $[0, 255]$ ADU. This range depends on the particular camera being used, in our case a high-speed 8-bit camera (Allied Vision Mako G030) synchronized with the swept laser source to capture interferograms corresponding to the evenly spaced $[k_1, \ldots, k_4] = [7.22, 7.36, 7.50, 7.63]$ rad/$\SI{}{\micro\meter}$. These particular values were chosen to be consistent with \cite{Chen2:16}, but any evenly spaced $k_n$ can be used. 

To generate our training and test sets, we begin by choosing a range of values for the parameters $[\alpha, V, L]$ that reflect typical values encountered in experiment. In our case, the completely unconstrained parameter sets are $\alpha \in [0, 255]$, $V\in [0, 1]$, and $L\in [0, \infty)$. For each fixed set of $\alpha$, $V$ and $L$, noise-free (mean) intensity values are generated according to Eq. \eqref{eq:int1}. These mean values are then converted to the average number of photo-electrons captured during camera exposure, $s_n = g \Bar{I}_n$, where $g = 34.4~e^-/\textrm{ADU}$ for our camera. The $s_n$ is the rate for the Poisson detection process, which we sample to obtain the actual numbers of photo-electrons (with noise) $x_n$, e.g. $\operatorname{Po(x_n; s_n})$. Finally, the noisy $I_n$ are obtained from $x_n$ through $I_n = \operatorname{int}(x_n / g)$. This is repeated for each of the four $k_n$ to generate one vector, $\mathbf{I}(m,n)$.

Further, since it is possible for the same OPL to have different $(\alpha, V)$, we randomly generate $K$ pairs of $(\alpha, V)$ for each OPL to ensure that the training data reflects the underlying physical process.

In general, for optimal sensitivity, we have found that the range of OPLs trained on for each network must satisfy $L_{\mathrm{max}} - L_{\mathrm{min}} \leq  2\pi/\operatorname{max(k_n)} \approx 0.8 ~\SI{}{\micro \meter}$. This guarantees that the training data only contains at most one period of the $I_n$ corresponding to $\operatorname{max(k_n)}$ (less than one period of the other $I_n$), but also limits the possible output values for each DNN. In practice, this is not a strict limitation as many biological samples have an OPL variation within $0.8$ $\SI{}{\micro \meter}$. In the case where larger OPL changes do occur, an adjacent neural network can be applied.

To summarize, if we select a range of OPLs of length $P$ to train on, our truth data is simply $\mathbf{L} = \left[L_1, L_2, \ldots, L_P  \right]$. These OPLs are then used to generate $K$ 4-vectors (the $\bar{I}_n$) for each OPL in $\mathbf{L}$. Therefore, our final training set size is $KP$, and the DNN's job during training is to associate a given intensity vector with the correct OPL. For all of the networks discussed in this work, we have used $K=P=1000$. 

\subsection{Simulation - Achieving the CRB}
\label{sse:simResults}

\begin{figure}[!ht]
    \centering
    \includegraphics[width=\linewidth]{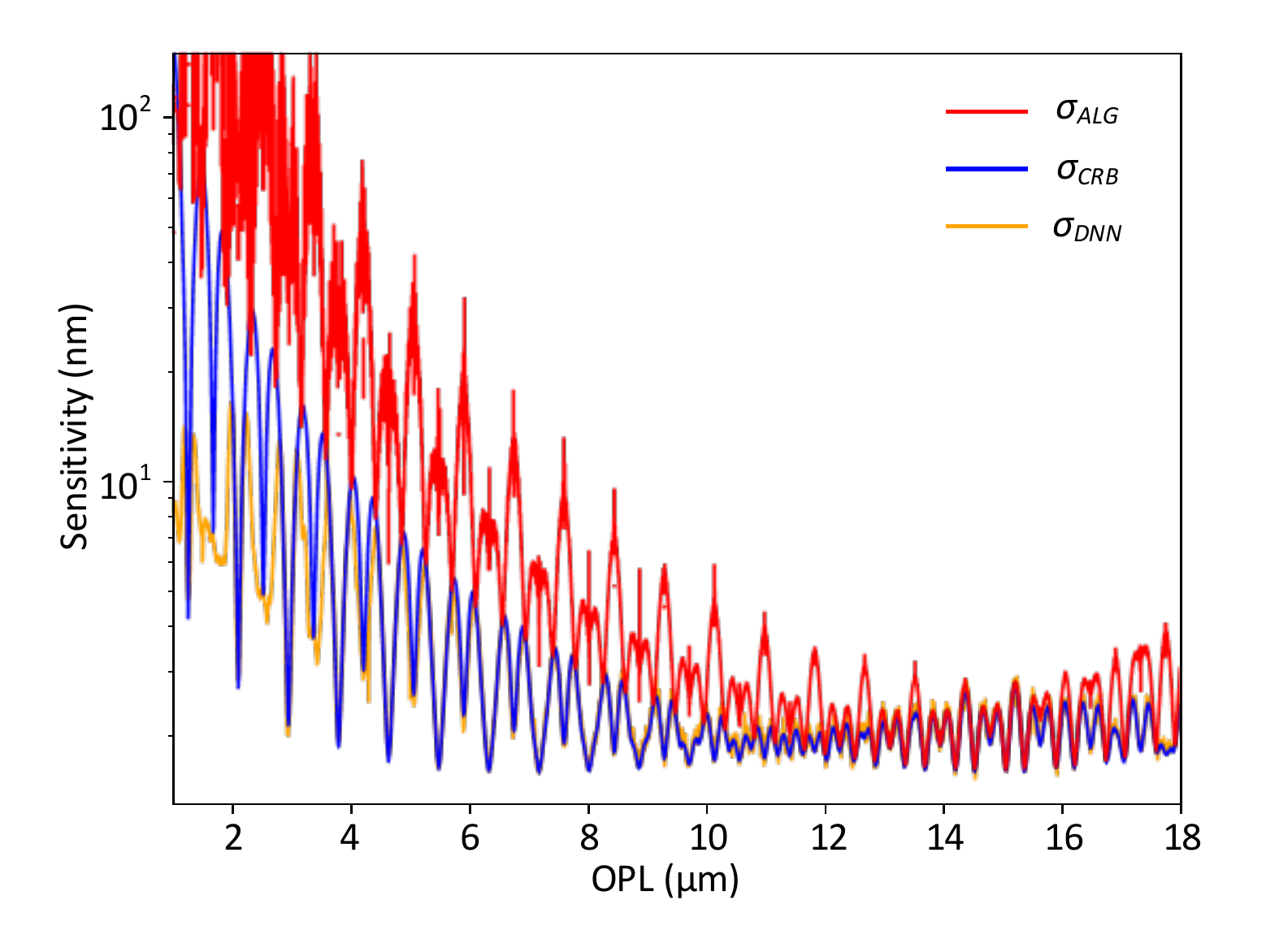}
    \caption{Sensitivity vs. OPL  plot including the DNN sensitivity. The blue ($\sigma_{CRB}$) and orange ($\sigma_{DNN}$) traces are nearly identical, suggesting the DNN achieves the lower bound set by Eq. \eqref{eq:crbDef}. }
    \label{fig:simResults1}
\end{figure}

In this section, we construct particular training sets and determine the temporal sensitivity of our DNNs trained with the inclusion of Poisson shot noise. In each case the networks were trained and then tested on a new data set to judge the performance. The output of the networks, $L_{DNN}$, is then recorded. To calculate a network's sensitivity, a Monte Carlo simulation was performed with $\alpha, V$ fixed, and the resulting intensity vector $\mathbf{I}$ were passed into the network $10, 000$ times with shot noise simulated as described previously. The standard deviation, $\sigma_{DNN} = \sqrt{\operatorname{Var}(L_{DNN}})$, of the resulting $10,000$ OPLs output by the DNN was then taken. This process is repeated for each OPL in the test set. The particular training set used in this case has $\alpha \in [70, 140]$ and $V \in [0.59, 0.95]$. The results are shown in Fig. \ref{fig:simResults1}.

\begin{figure}[!ht]
    \centering
    \includegraphics[width=\linewidth]{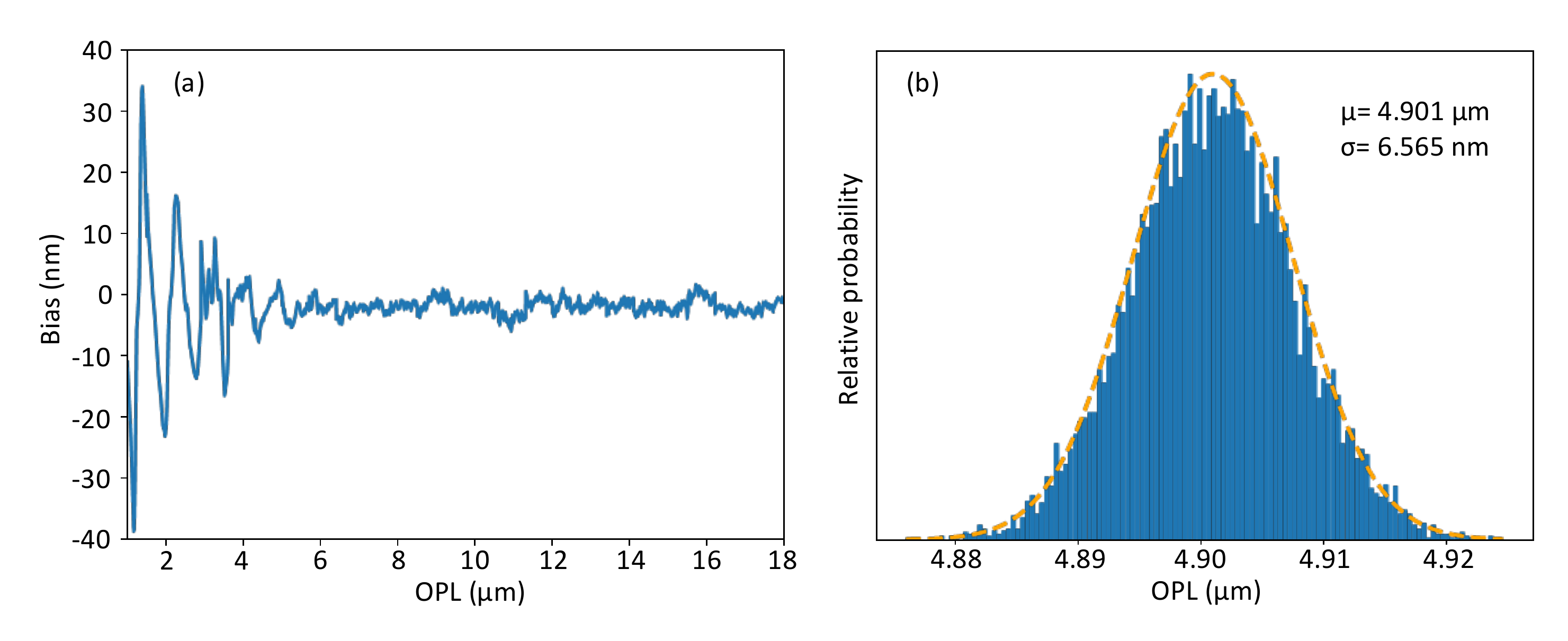}
    \caption{(a) Bias (average error) of the neural network vs. OPL. Across much of the range, the neural network is highly accurate ($< 5$ nm absolute mean error). (b) Histogram of DNN output for a true OPL in the test set of $4.901$ $\SI{}{\micro\meter}$. The sensitivity for this OPL is $6.565$ nm.}
    \label{fig:histo1}
\end{figure}

Figure \ref{fig:histo1} shows the bias and statistics of the outputs from the DNNs trained with noise. The bias shows the average deviation of the DNN output from true OPL. For each OPL, its value is well below the corresponding sensitivity, suggesting the bias error is negligible. Figure \hyperref[fig:histo1]{5(b)} shows the distribution of DNN output for a fixed OPL with an expected Gaussian distribution (approximate of Poisson shot noise). The sensitivity is simply the standard deviation of the distribution and precisely corresponds to the sensitivity along the $\sigma_{DNN}$ curve in Fig. \ref{fig:simResults1} for an OPL difference of 4.901 $\SI{}{\micro \meter}$.
To generate Figs. 4 and 5 over the entire OPL range of 1-18 $\SI{}{\micro \meter}$, multiple neural networks were trained on smaller but overlapping ranges for $L$ (with the same ranges for $\alpha, V$), each with width $L_{\mathrm{max}} - L_{\mathrm{min}} =  0.8$ $\SI{}{\micro\meter}$ as discussed in Sec.  \hyperref[subSec:TDG]{(3.2)}. This is done to avoid periodic ambiguities within the intensity training data, similar to phase wrapping. 

Note that $\sigma_{DNN}$ closely follows the CRB, and outperforms the Carr\'e approach (ALG) across much of the range. This confirms that DNNs informed with noise statistics can achieve CRB. Of particular interest are the sensitivities obtained by the DNN at low OPLs, where the DNN begins to \emph{exceed} $\sigma_{CRB}$. As we show in Sec. \hyperref[sse:exCRB]{(3.4)}, by constraining the range of $\alpha$ and $V$, we are informing the DNNs with new knowledge. Therefore, $\sigma_{CRB}$ no longer represents the fundamental limit for this DNN.  

\subsection{Simulation - Exceeding the CRB with Parameter Constraints}
\label{sse:exCRB}

\begin{figure}[!ht]
    \centering
    \includegraphics[width=\linewidth]{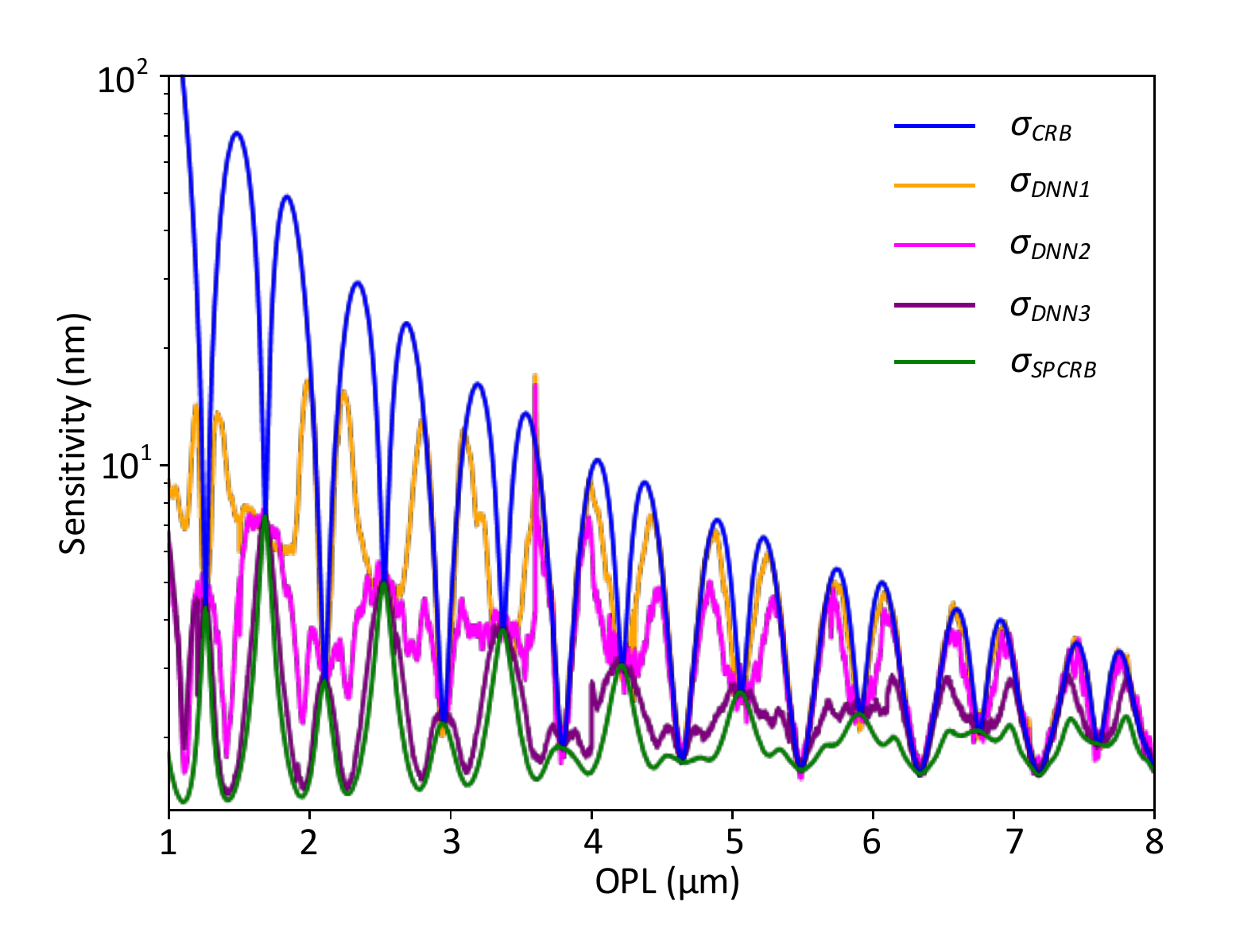}
    \caption{Sensitivities of DNNs trained on different ranges for $\alpha$ and $V$. $\sigma_{CRB}$ and $\sigma_{SPCRB}$ are plotted for reference. OPL range shown is shortened to emphasize the differences between the networks. Note that $\sigma_{DNN3}$, trained on the most confined range, approaches the bound given in Eq. \eqref{eq:spcrbDef}.}
    \label{fig:simResults2}
\end{figure}

In optical interferometry, the values of $\alpha$ and $V$ can almost always be well estimated with good precision. For example, the intensities of sample arm and reference arms, as well as interference efficiency can be measured. In WSI, they can even be calculated from $I_1$-$I_4$. In other words, the uncertainty intervals of $\alpha$ and $V$ can be made rather narrow. Such additional information, if incorporated into the DNN training together with noise statistics, would further improve the sensitivity to reach well beyond the traditional CRB, therefore achieving CCRB.

To verify this, we trained three different sets of DNNs using progressively narrower ranges for $\alpha$ and $V$, representing increasing knowledge about them.  The results are shown in Fig. \ref{fig:simResults2}, where $\sigma_{DNN1}$ corresponds to $\alpha \in [70, 140]$ ADU, $V\in[0.59, 0.95]$, $\sigma_{DNN2}$ corresponds to $\alpha \in [115, 135]$ ADU, $V\in [0.65, 0.75]$, and $\sigma_{DNN3}$ was trained with $\alpha \in [125, 130]$ ADU, $V\in [0.67, 0.72]$. The subscripts 1,2,3 refer to the specific DNNs used in this simulation.  The broad range of values for $\sigma_{DNN1}$ is chosen to ensure accommodation of all typical sample conditions for our WSI experiments with abundant margin. Therefore it mostly matches CRB with some advantages only for low OPL values. In comparison, the range chosen for $\sigma_{DNN2}$ is smaller and thus $ \sigma_{DNN1} > \sigma_{DNN2}$. This generally corresponds to the cases where DNNs can be better tailored to a specific sample. Lastly, the even narrower range chosen for $\sigma_{DNN3}$ produces more significant increase in sensitivity that approaches the limit of the SPCRB. To gain such benefits, one may need to segment the image into small, sufficiently uniform areas and train DNNs specifically for each area, or simply process each point individually. Although this increases the processing load, it is well justified because of the tremendous gain in quantification sensitivity and SNR. Summarizing, it is clear from Fig. \ref{fig:simResults2} that:
\begin{equation}
\label{eq:inEq2}
\sigma_{ALG}  \geq \sigma_{CRB} \geq \sigma_{DNN1} \geq \sigma_{DNN2} \geq  \sigma_{DNN3} \geq \sigma_{SPCRB}
\end{equation}
This suggests that by using a practical range for $\alpha$ and $V$ when training the DNN, we can indeed enter the regime of CCRB, not only achieving the unconstrained CRB given by Eq. \eqref{eq:crbDef}, but surpassing it and beginning to approach the SPCRB in Eq. \eqref{eq:spcrbDef}. No changes need to be made to the original optical system to take advantage of the improvement in sensitivity provided by the DNNs. Next, to further validate these findings, we experimentally calculate the sensitivities of our DNNs and show that our simulations are accurate.

\section{Experimental demonstration}
\label{sec:exp}

\begin{figure}[!ht]
    \centering
    \includegraphics[width=\linewidth]{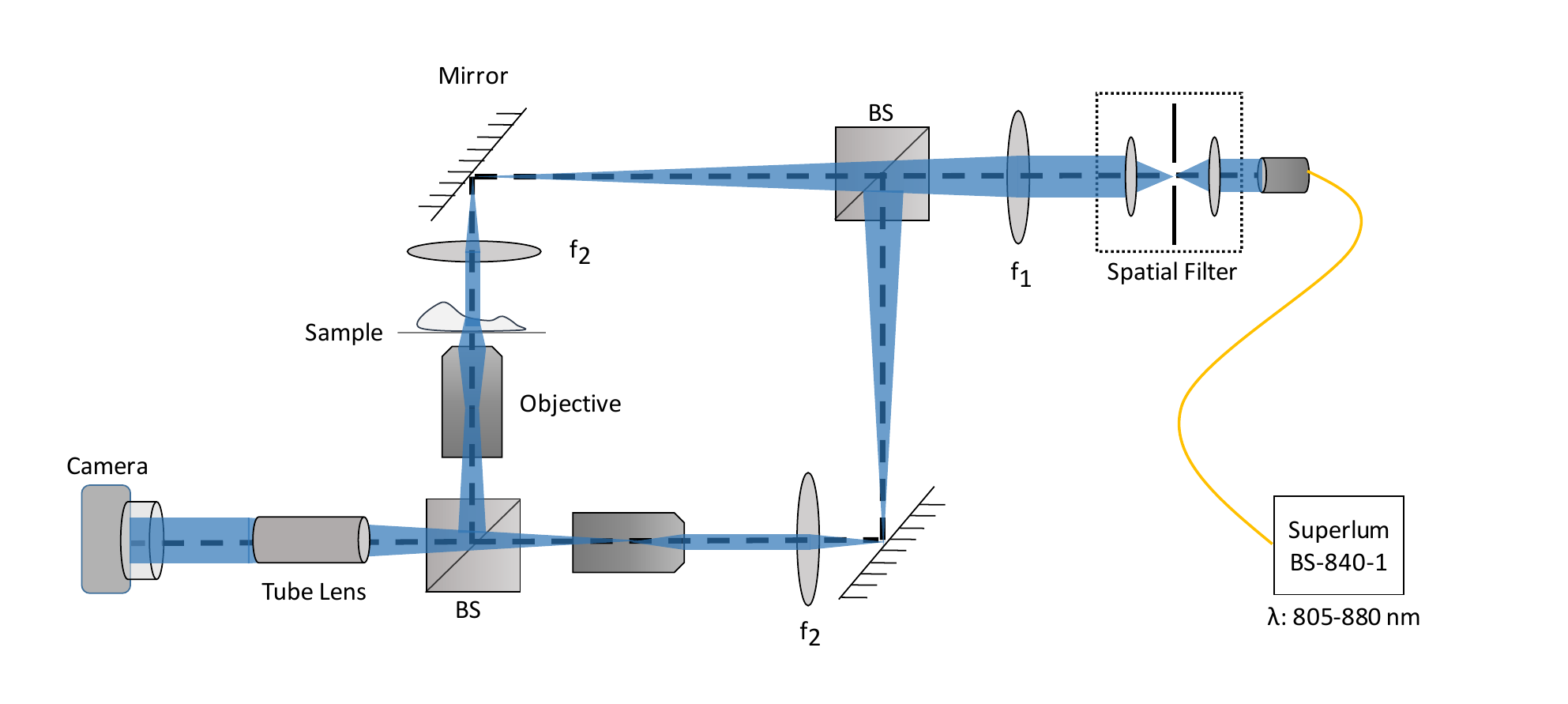}
    \caption{A typical WSI system consisting of a Mach-Zehnder interferometer containing two identical beamsplitters (BS), objectives, and mirrors. Our swept source laser is the Superlum BS-840-1 which can sweep from 805-880 $\mathrm{nm}$.}
    \label{fig:sysDiagram}
\end{figure}

\begin{figure*}
    \centering
    \includegraphics[scale=0.375]{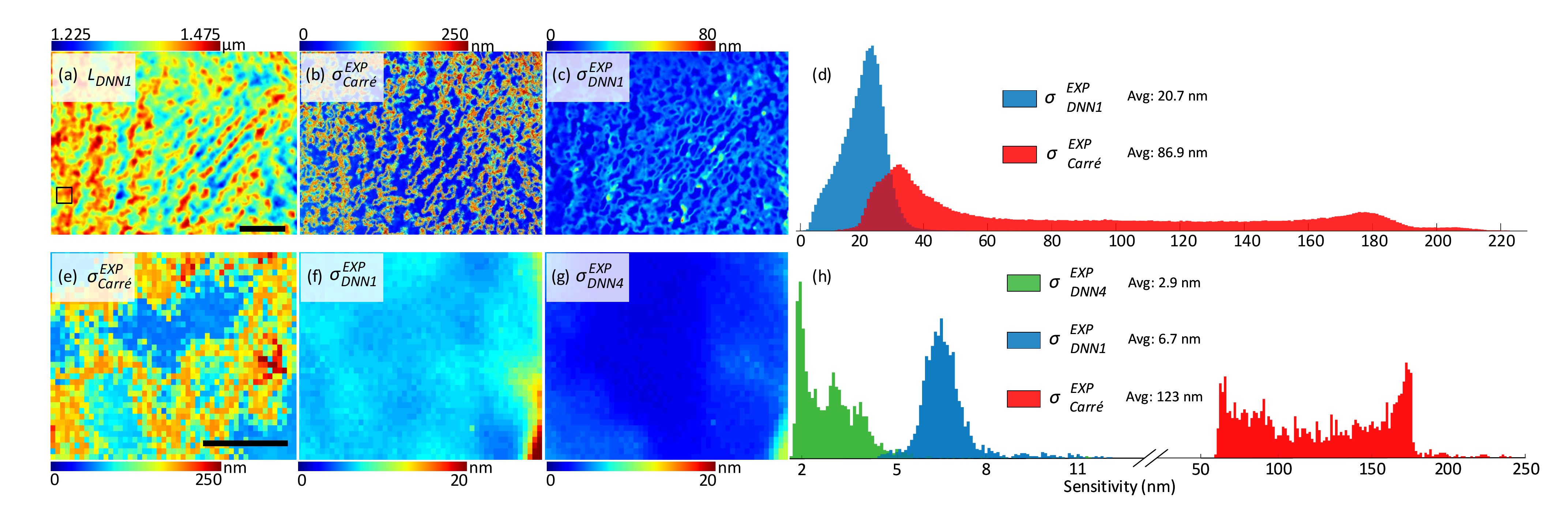}
    \caption{(a) Averaged OPL output of $\mathrm{DNN_1}$. The small rectangle corresponds to the selected region of interest ($6.6~ \SI{}{\micro \meter}  \times  9.4 ~\SI{}{\micro \meter}$) in (e)-(h). (b), (c) Sensitivity maps of $\sigma^{EXP}_{Carr\acute{e}}$ and $\sigma^{EXP}_{DNN1}$. (d) Histogram comparing $\sigma^{EXP}_{DNN1}$ and $\sigma^{EXP}_{Carr\acute{e}}$. $DNN_1$ shows vast improvement across the full field of view when compared with Carr\'e. Scale bar for (a)-(c): 20 $\SI{}{\micro \meter}$. (e), (f) Selected region processed by $\sigma^{EXP}_{Carr\acute{e}}$ and $\sigma^{EXP}_{DNN1}$. (g) Selected region of $\sigma^{EXP}_{DNN4}$, which was obtained by training a DNN based on the $\alpha$ and $V$ information determined from our experimental data. Scale bar for (e)-(g): 3 $\SI{}{\micro \meter}$. (h) Histogram comparing $\sigma^{EXP}_{DNN1}$, $\sigma^{EXP}_{DNN4}$, and $\sigma^{EXP}_{Carr\acute{e}}$. We again observe a marked improvement by informing the training set for $\mathrm{DNN_4}$ with the approximate ranges of $\alpha$ and $V$ determined from average OPL distribution, $L_{DNN1}$.}
    \label{fig:sensExp1}
\end{figure*}

\subsection{Sensitivity}
\label{subsec:expSens}
To test our DNNs experimentally, we make use of the WSI system illustrated in Fig. \ref{fig:sysDiagram}, where a swept laser source (Superlum BS-840-1, 805-880 nm) is spatially filtered and collimated into a traditional Mach-Zehnder interferometer. Our first experiment consists of a blank sample (a glass cover slip) inserted into the sample arm with $L=L_S - L_R$ adjusted to a particular value. As the first example, we tuned the OPL difference between the two arms of the interferometer to an average value $L \approx 1.4$ $\SI{}{\micro \meter}$ across the field of view. We then made consecutive acquisitions to produce 500 phase images via our $\mathrm{DNN_1}$ and Eq. \eqref{eq:carre}. For clarity, we refer to the experimentally obtained sensitivities as $(\cdot)^{EXP}$. For example, the sensitivity associated with a DNN applied to experimental data is $(\sigma^{EXP}_{DNN})$. 

For $L\approx 1.4 ~\SI{}{\micro \meter}$, we expect that the Carr\'e algorithm Eq. \eqref{eq:carre} will struggle to provide an accurate demodulation as the predicted sensitivity for this OPL is in the hundreds of nanometers according to Fig. \ref{fig:crb_alg_spcrb}. On the other hand, $\mathrm{DNN_1}$ retain sensitivities around 10 nanometers or better (see Fig. \ref{fig:simResults2}). Thus, we expect $\mathrm{DNN_1}$ to be able to demodulate the raw data at this $L$ with much greater sensitivity. The results are shown in Figs. \hyperref[fig:sensExp1]{8(a)}-\hyperref[fig:sensExp1]{8(d)} , where sensitivity and image quality are greatly improved from $(\sigma^{EXP}_{Carr\acute{e}})$ to $(\sigma^{EXP}_{DNN1})$. In Fig. \hyperref[fig:sensExp1]{8(d)}, the mean sensitivity for Carr\'e algorithm is 86.9 nm, while the mean sensitivity for $\mathrm{DNN_1}$ is 20.7 nm, making $\mathrm{DNN_1}$ 4.2 times more sensitive than Eq. \eqref{eq:carre} which represents an SNR increase of ~12 dB.

Having established this initial comparison, we selected a small region ( $6.6~ \SI{}{\micro \meter}  \times  9.4~ \SI{}{\micro \meter}$) within the field over which the ranges of values which $\alpha$ and $V$  take on was reduced. To take advantage of this tighter constraints, we trained another neural network, $\mathrm{DNN_4}$, with $\alpha\in [60, 90]$ and $V\in[0.8, 1]$. These values were chosen for the training set to ensure $\mathrm{DNN_4}$ will demodulate this data with greater sensitivity. The results are shown in Figs. \hyperref[fig:sensExp1]{8(e)}-\hyperref[fig:sensExp1]{8(h)}, where the sensitivity distributions $\sigma^{EXP}_{Carr\acute{e}}$, $\sigma^{EXP}_{DNN1}$, and $\sigma^{EXP}_{DNN4}$ for this cropped region are shown and then plotted together as a histogram. Fig. \hyperref[fig:sensExp1]{8(h)} quantitatively verifies that $\mathrm{DNN_4}$ possesses the best sensitivity distribution. Across the selected area, the average values of the sensitivity for these three processing methods are 123 nm, 6.7 nm, and 2.9 nm. $\mathrm{DNN_1}$ is 18 times better than Carr\'e, equivalent to an SNR gain of ~25 dB, while $\mathrm{DNN_4}$ is 42 times better with an SNR gain of ~33 dB. All together, these results provide experimental support for the inequality in Eq. \eqref{eq:inEq2}. Additionally, they confirm that the DNNs are capable of providing a substantially more efficient OPL demodulation when Eq. \eqref{eq:carre} fails (e.g. when the argument of the square root is negative).

With the improvement confirmed, it is also important to know whether the experimental results follow the simulated results in Fig. \ref{fig:simResults2} accurately. That is, do the DNNs have the same $\sigma$ vs. $L$ relationships when applied to experimental data? Here we take advantage of a small OPL variation across the cropped region of the blank sample indicated in Fig. \hyperref[fig:sensExp1]{8(a)}. This allows us to take a line from the smaller image (e.g. the center line) and plot the DNNs temporal sensitivity against their average OPL output across the 500 acquisitions to obtain an experimental plot analogous to Fig. \ref{fig:simResults2}, albeit for a narrow range of OPL differences. \\
\begin{figure}[!ht]
    \centering
    \includegraphics[width=\linewidth]{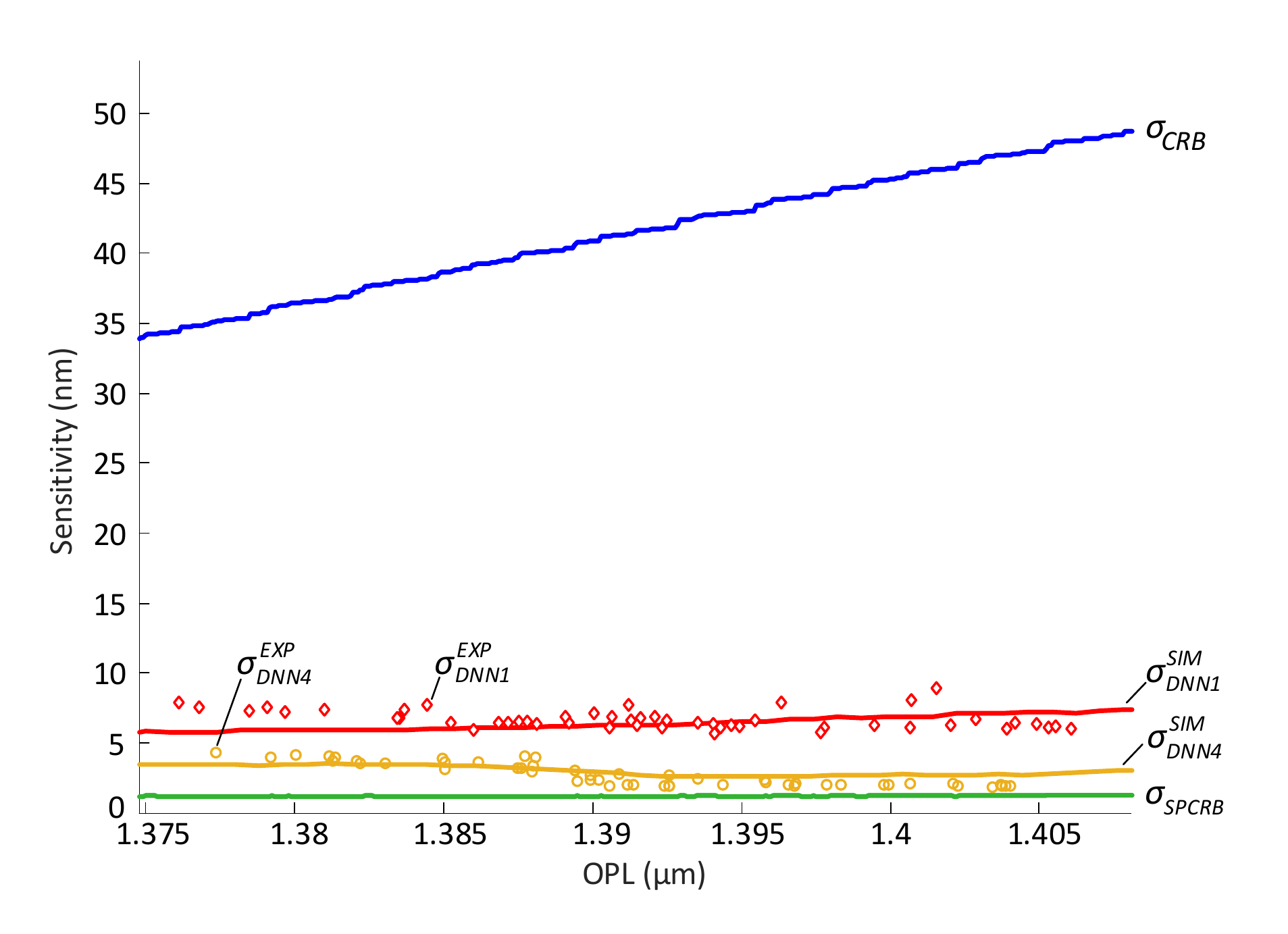}
    \caption{Experimental and simulated sensitivities for $\mathrm{DNN_1}$ and $\mathrm{DNN_4}$. $\sigma_{CRB}$ and $\sigma_{SPCRB}$ are also shown. Note that both DNN sensitivities shown here surpass $\sigma_{CRB}$ for much of the range shown, and closely follow the predicted values indicated by the solid curves for $\sigma^{SIM}_{DNN1}$ and $\sigma^{SIM}_{DNN4}$.  The values $\alpha = 75$ ADU and $V=0.93$ were used to produce the simulated curves. These values correspond to the average value of these parameters across the center line of the cropped image in Fig. \hyperref[fig:sensExp1]{8(e)-(g)}. }
    \label{fig:sensExpCrbSpCrb}
\end{figure}
~~~The result is shown in Fig.$~$\ref{fig:sensExpCrbSpCrb}, which demonstrates that $\mathrm{DNN_1}$ and $\mathrm{DNN_4}$ follow the CCRB values predicted by the simulation accurately, and are located between CRB and SPCRB, as expected. These results show that it is indeed possible to use DNNs for informed OPL (phase) demodulation to surpass the traditional CRB given by Eq. \eqref{eq:crbDef} by making use of parametric constraints for $\alpha$ and $V$ in Eq. \eqref{eq:int1}. In particular, $(\sigma^{EXP}_{DNN1})$ exceeds $\sigma_{CRB}$ for much of the range shown in Fig. \ref{fig:sensExpCrbSpCrb}, despite being trained on broad parameter ranges, and further sensitivity gains are seen with $(\sigma^{EXP}_{DNN4})$ by virtue  of being trained on smaller ranges. Across this range of OPL differences (about 1.37 to 1.4 $\SI{}{\micro \meter}$), the sensitivity improvement for $(\sigma^{EXP}_{DNN4})$ exceeds an order of magnitude as compared with $\sigma_{CRB}$. That such a large enhancement factor can be attained by simply altering the training set to reflect the well-conditioned WSI system is remarkable.

\subsection{Live Cell Imaging}
\label{subsec:expCells}

\begin{figure}[!ht]
    \centering
    \includegraphics[width=\linewidth]{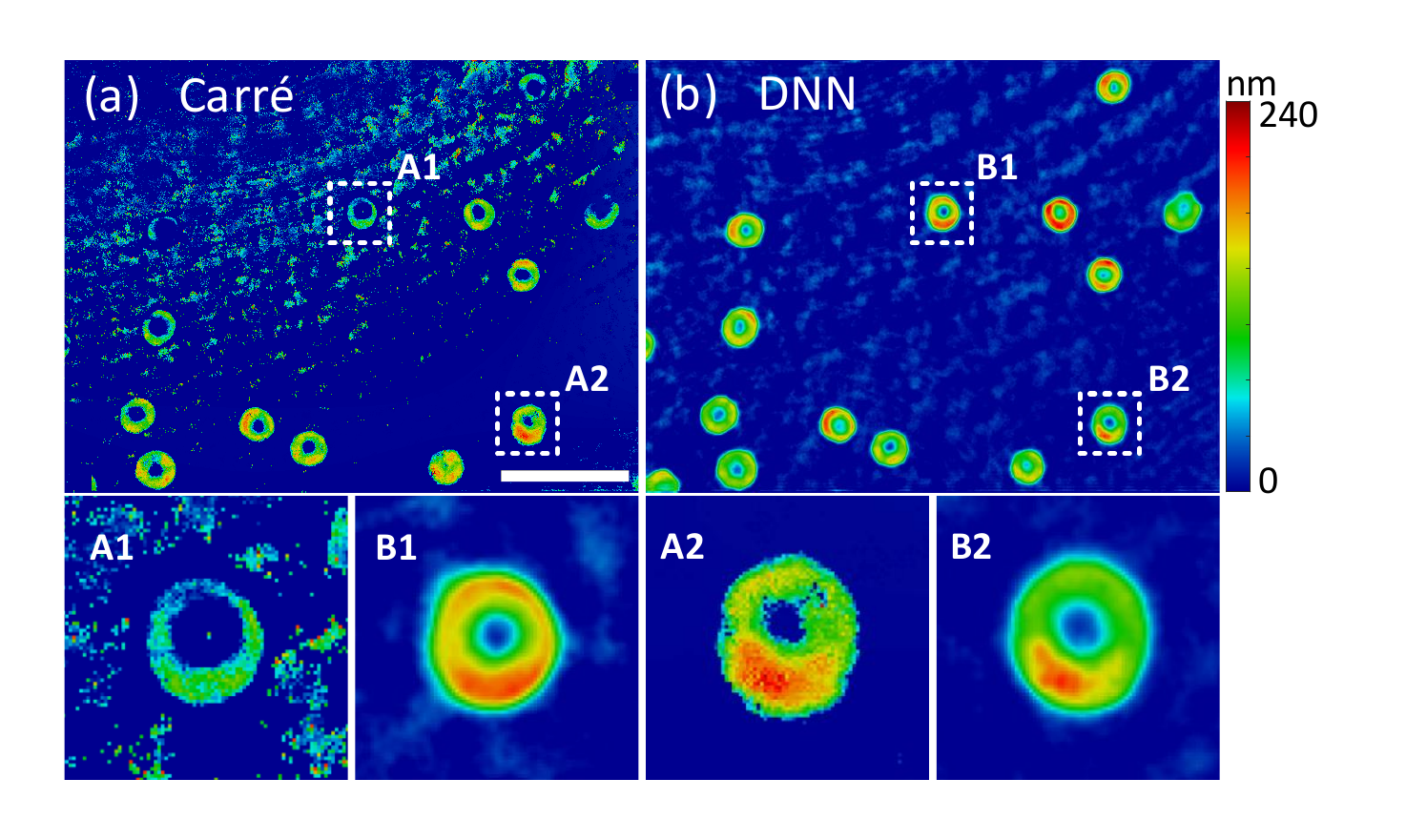}
    \caption{(a) RBC image generated by Eq. \eqref{eq:carre}. Note the phase artifacts and how some of the RBCs visible in the DNN image are either difficult to see or completely missing in this image. (b) RBC image produced by the DNN, showing a relatively smooth background with well defined RBC. Scale bar for (a) and (b): 20 $\SI{}{\micro \meter}$. Bottom: Zoomed in comparisons between the two approaches for the cells indicated in (a) and (b).}
    \label{fig:newRBC}
\end{figure}

We imaged human red blood cells (RBCs) to further validate the use of DNN processing with live cell data, and to demonstrate the advantage of the DNNs over Eq. \eqref{eq:carre}. The results are shown in Fig. \ref{fig:newRBC}. For this experiment, we have made use of a DNN trained with $\alpha \in [70, 140]$ ADU, $V \in [0.59, 0.95]$ (the same as $\mathrm{DNN_1}$) as this is the most flexible in terms of demodulating the entire field of view of our camera. Additionally, we have set $L_0 \approx 7~ \SI{}{\micro\meter}$, and as such the DNN used to generate Fig. \hyperref[fig:newRBC]{10(b)} was trained on simulated data corresponding to an OPL range which contains $7~ \SI{}{\micro\meter}$ (e.g. $L\in [6.7, 7.5] \SI{}{\micro\meter}$). By doing so, we can ensure that $L(x,y)$ across the entire image will lie within the range which the DNN was trained on. In this region, based on  Fig. \ref{fig:simResults1}, we can see that the DNN has a distinct advantage over the Carr\'e equation, as confirmed by Fig. \ref{fig:newRBC}. The phase image produced by the DNN is clear and free of artifacts, whereas the image produced by Eq. \eqref{eq:carre} contains many artifacts and subsequently some of the RBCs visible in the DNN image are missing in the Carr\'e image. These findings are consistent with the predictions of our simulations, and suggest that the DNNs trained on purely simulated data can produce excellent phase images from raw intensity data obtained from an experimental system, provided that such system and its noise statistics can be well modelled. In practice, shot noise-limited operation can often be achieved and is thus a good place to apply DNN processing. 

\section{Discussion and Conclusions}
\label{sec:conclusions}

In optical sensing and imaging, one often has a certain amount of knowledge about the sample and/or the measurement process that is not utilized in traditional signal processing. Noise statistics and parameter bounds are examples of such information. Neither is considered in Eq. (2). In this work, we have demonstrated that such information can be conveniently incorporated into DNN-based signal processing with the benefit of substantially improved sensitivity and SNR. In some cases, such improvements are remarkable. Because of the square relationship between OPL (phase) sensitivity and SNR, a linear increase in sensitivity is equivalent to a power of 2 increase in SNR, which is much more efficient than directly increasing SNR.

We based our study on an expanded 5-tier framework of sensitivity evaluation, with the newly introduced CCRB and SPCRB. The framework is essential in elucidating the relationship between key sensitivity metrics. It also makes clear the roles of system information, such as noise statistics and parameter constraints, and how they can be used to approach and exceed sensitivity limits.  Specifically, we followed an informed, physical model-based approach. First, we train our networks with shot noise statistics, and were able to exceed ALG and achieve CRB. Then the parametric constraints allow us to exceed CRB, achieve CCRB and approach SPCRB.

It is also demonstrated that DNNs trained with simulated data are applicable for processing experimental data to gain expected improvement under practical conditions. It is important that the physical process of the system can be well modelled, which is common in many areas of the quantitative sensing and imaging.

Such sensitivity improvement will not only improve data quality, but may be used to benefit system hardware implementation. For example, when significant sensitivity and SNR gain is available, it may be traded for relaxing hardware requirements, e.g. lower light power, shorter camera exposure, higher speed, and potentially lower cost implementations, which may eventually enable low-cost yet high-performance optical metrology systems.

While we demonstrate DNN processing using WSI, the same concept may be similarly applied to other interferometry techniques or other measurement systems in general. Each case will be different due to their different models and underlying physics. Even for WSI, if the wavenumbers are unevenly spaced, the sensitivity curves will be different. These will be the subject of our future research. Nonetheless, we believe the concept remains valid that injecting new knowledge into signal processing would improve measurement sensitivity, and that DNNs represent a valuable tool for demodulating and quantifying optical signals.

\section*{Acknowledgments.}
J.A.B. and J.G.T. gratefully acknowledge the support of Bradley Fellowship from the Bradley Department of Electrical and Computer Engineering at Virginia Tech.

\bibliographystyle{elsarticle-num-names} 
\bibliography{myBib}

\begin{thebibliography}{28}
\expandafter\ifx\csname natexlab\endcsname\relax\def\natexlab#1{#1}\fi
\providecommand{\url}[1]{\texttt{#1}}
\providecommand{\href}[2]{#2}
\providecommand{\path}[1]{#1}
\providecommand{\DOIprefix}{doi:}
\providecommand{\ArXivprefix}{arXiv:}
\providecommand{\URLprefix}{URL: }
\providecommand{\Pubmedprefix}{pmid:}
\providecommand{\doi}[1]{\href{http://dx.doi.org/#1}{\path{#1}}}
\providecommand{\Pubmed}[1]{\href{pmid:#1}{\path{#1}}}
\providecommand{\bibinfo}[2]{#2}
\ifx\xfnm\relax \def\xfnm[#1]{\unskip,\space#1}\fi
\bibitem[{Popescu(2011)}]{pope1}
\bibinfo{author}{G.~Popescu}, \bibinfo{title}{Quantitative Phase Imaging of
  Cells and Tissues}, \bibinfo{year}{2011}.
\bibitem[{Cuche et~al.(1999)Cuche, Bevilacqua, and Depeursinge}]{Cuche:99}
\bibinfo{author}{E.~Cuche}, \bibinfo{author}{F.~Bevilacqua},
  \bibinfo{author}{C.~Depeursinge},
\newblock \bibinfo{title}{Digital holography for quantitative phase-contrast
  imaging},
\newblock \bibinfo{journal}{Opt. Lett.} \bibinfo{volume}{24}
  (\bibinfo{year}{1999}) \bibinfo{pages}{291--293}. \URLprefix
  \url{http://ol.osa.org/abstract.cfm?URI=ol-24-5-291}.
  \DOIprefix\doi{10.1364/OL.24.000291}.
\bibitem[{Wang et~al.(2011)Wang, Millet, Mir, Ding, Unarunotai, Rogers,
  Gillette, and Popescu}]{Wang:11}
\bibinfo{author}{Z.~Wang}, \bibinfo{author}{L.~Millet},
  \bibinfo{author}{M.~Mir}, \bibinfo{author}{H.~Ding},
  \bibinfo{author}{S.~Unarunotai}, \bibinfo{author}{J.~Rogers},
  \bibinfo{author}{M.~U. Gillette}, \bibinfo{author}{G.~Popescu},
\newblock \bibinfo{title}{Spatial light interference microscopy (slim)},
\newblock \bibinfo{journal}{Opt. Express} \bibinfo{volume}{19}
  (\bibinfo{year}{2011}) \bibinfo{pages}{1016--1026}. \URLprefix
  \url{http://www.opticsexpress.org/abstract.cfm?URI=oe-19-2-1016}.
  \DOIprefix\doi{10.1364/OE.19.001016}.
\bibitem[{Gillies et~al.(2018)Gillies, Gamal, Downes, Reinwald, Yang, Haj, and
  Bagnaninchi}]{GILLIES2018126}
\bibinfo{author}{D.~Gillies}, \bibinfo{author}{W.~Gamal},
  \bibinfo{author}{A.~Downes}, \bibinfo{author}{Y.~Reinwald},
  \bibinfo{author}{Y.~Yang}, \bibinfo{author}{A.~E. Haj},
  \bibinfo{author}{P.~Bagnaninchi},
\newblock \bibinfo{title}{Real-time and non-invasive measurements of cell
  mechanical behaviour with optical coherence phase microscopy},
\newblock \bibinfo{journal}{Methods} \bibinfo{volume}{136}
  (\bibinfo{year}{2018}) \bibinfo{pages}{126 -- 133}. \URLprefix
  \url{http://www.sciencedirect.com/science/article/pii/S1046202317301809}.
  \DOIprefix\doi{https://doi.org/10.1016/j.ymeth.2017.10.010},
  \bibinfo{note}{methods in Quantitative Phase Imaging in Life Science}.
\bibitem[{Kim et~al.(2018)Kim, Oh, Kim, Lee, Pack, Park, and Park}]{KIM2018160}
\bibinfo{author}{D.~Kim}, \bibinfo{author}{N.~Oh}, \bibinfo{author}{K.~Kim},
  \bibinfo{author}{S.~Lee}, \bibinfo{author}{C.-G. Pack},
  \bibinfo{author}{J.-H. Park}, \bibinfo{author}{Y.~Park},
\newblock \bibinfo{title}{Label-free high-resolution 3-d imaging of gold
  nanoparticles inside live cells using optical diffraction tomography},
\newblock \bibinfo{journal}{Methods} \bibinfo{volume}{136}
  (\bibinfo{year}{2018}) \bibinfo{pages}{160 -- 167}. \URLprefix
  \url{http://www.sciencedirect.com/science/article/pii/S1046202317301792}.
  \DOIprefix\doi{https://doi.org/10.1016/j.ymeth.2017.07.008},
  \bibinfo{note}{methods in Quantitative Phase Imaging in Life Science}.
\bibitem[{Nadeau et~al.(2018)Nadeau, Park, and Popescu}]{NADEAU20181}
\bibinfo{author}{J.~Nadeau}, \bibinfo{author}{Y.~Park},
  \bibinfo{author}{G.~Popescu},
\newblock \bibinfo{title}{Methods in quantitative phase imaging in life
  science},
\newblock \bibinfo{journal}{Methods} \bibinfo{volume}{136}
  (\bibinfo{year}{2018}) \bibinfo{pages}{1 -- 3}. \URLprefix
  \url{http://www.sciencedirect.com/science/article/pii/S1046202318300744}.
  \DOIprefix\doi{https://doi.org/10.1016/j.ymeth.2018.03.004},
  \bibinfo{note}{methods in Quantitative Phase Imaging in Life Science}.
\bibitem[{{Jo} et~al.(2019){Jo}, {Cho}, {Lee}, {Choi}, {Kim}, {Min}, and
  {Park}}]{qpiML}
\bibinfo{author}{Y.~{Jo}}, \bibinfo{author}{H.~{Cho}}, \bibinfo{author}{S.~Y.
  {Lee}}, \bibinfo{author}{G.~{Choi}}, \bibinfo{author}{G.~{Kim}},
  \bibinfo{author}{H.~{Min}}, \bibinfo{author}{Y.~{Park}},
\newblock \bibinfo{title}{Quantitative phase imaging and artificial
  intelligence: A review},
\newblock \bibinfo{journal}{IEEE Journal of Selected Topics in Quantum
  Electronics} \bibinfo{volume}{25} (\bibinfo{year}{2019})
  \bibinfo{pages}{1--14}. \DOIprefix\doi{10.1109/JSTQE.2018.2859234}.
\bibitem[{{Rivenson} et~al.(2019){Rivenson}, {Liu}, {Wei}, {De Haan}, {Zhan},
  and {Ozcan}}]{phaseStain}
\bibinfo{author}{Y.~{Rivenson}}, \bibinfo{author}{T.~{Liu}},
  \bibinfo{author}{Z.~{Wei}}, \bibinfo{author}{K.~{De Haan}},
  \bibinfo{author}{Y.~{Zhan}}, \bibinfo{author}{A.~{Ozcan}},
\newblock \bibinfo{title}{Phasestain: Deep learning-based histological staining
  of quantitative phase images},
\newblock in: \bibinfo{booktitle}{2019 Conference on Lasers and Electro-Optics
  (CLEO)}, \bibinfo{year}{2019}, pp. \bibinfo{pages}{1--2}.
  \DOIprefix\doi{10.23919/CLEO.2019.8750359}.
\bibitem[{{Kellman} et~al.(2019){Kellman}, {Bostan}, {Repina}, and
  {Waller}}]{physicsDNN}
\bibinfo{author}{M.~R. {Kellman}}, \bibinfo{author}{E.~{Bostan}},
  \bibinfo{author}{N.~A. {Repina}}, \bibinfo{author}{L.~{Waller}},
\newblock \bibinfo{title}{Physics-based learned design: Optimized
  coded-illumination for quantitative phase imaging},
\newblock \bibinfo{journal}{IEEE Transactions on Computational Imaging}
  \bibinfo{volume}{5} (\bibinfo{year}{2019}) \bibinfo{pages}{344--353}.
  \DOIprefix\doi{10.1109/TCI.2019.2905434}.
\bibitem[{Rivenson et~al.(2017)Rivenson, Zhang, Gunaydin, Teng, and
  Ozcan}]{prHolo}
\bibinfo{author}{Y.~Rivenson}, \bibinfo{author}{Y.~Zhang},
  \bibinfo{author}{H.~Gunaydin}, \bibinfo{author}{D.~Teng},
  \bibinfo{author}{A.~Ozcan},
\newblock \bibinfo{title}{Phase recovery and holographic image reconstruction
  using deep learning in neural networks},
\newblock \bibinfo{journal}{https://arxiv.org/abs/1705.04286}
  \bibinfo{volume}{7} (\bibinfo{year}{2017}).
  \DOIprefix\doi{10.1038/lsa.2017.141}.
\bibitem[{Sinha et~al.(2017)Sinha, Lee, Li, and Barbastathis}]{Sinha:17}
\bibinfo{author}{A.~Sinha}, \bibinfo{author}{J.~Lee}, \bibinfo{author}{S.~Li},
  \bibinfo{author}{G.~Barbastathis},
\newblock \bibinfo{title}{Lensless computational imaging through deep
  learning},
\newblock \bibinfo{journal}{Optica} \bibinfo{volume}{4} (\bibinfo{year}{2017})
  \bibinfo{pages}{1117--1125}. \URLprefix
  \url{http://www.osapublishing.org/optica/abstract.cfm?URI=optica-4-9-1117}.
  \DOIprefix\doi{10.1364/OPTICA.4.001117}.
\bibitem[{Kemp(2018)}]{Kemp_2018}
\bibinfo{author}{Z.~D.~C. Kemp},
\newblock \bibinfo{title}{Propagation based phase retrieval of simulated
  intensity measurements using artificial neural networks},
\newblock \bibinfo{journal}{Journal of Optics} \bibinfo{volume}{20}
  (\bibinfo{year}{2018}) \bibinfo{pages}{045606}. \URLprefix
  \url{https://doi.org/10.1088%2F2040-8986%2Faab02f}.
  \DOIprefix\doi{10.1088/2040-8986/aab02f}.
\bibitem[{Wu et~al.(2019)Wu, Yang, Li, and Zhu}]{xRayTomog}
\bibinfo{author}{Z.~Wu}, \bibinfo{author}{T.~Yang}, \bibinfo{author}{L.~Li},
  \bibinfo{author}{Y.~Zhu}, \bibinfo{title}{Hierarchical convolutional network
  for sparse-view x-ray ct reconstruction}, \bibinfo{year}{2019}. \URLprefix
  \url{https://doi.org/10.1117/12.2521239}. \DOIprefix\doi{10.1117/12.2521239}.
\bibitem[{{Wang} et~al.(2019){Wang}, {Xie}, and {Zeng}}]{rpNET}
\bibinfo{author}{L.~{Wang}}, \bibinfo{author}{C.~{Xie}},
  \bibinfo{author}{N.~{Zeng}},
\newblock \bibinfo{title}{Rp-net: A 3d convolutional neural network for brain
  segmentation from magnetic resonance imaging},
\newblock \bibinfo{journal}{IEEE Access} \bibinfo{volume}{7}
  (\bibinfo{year}{2019}) \bibinfo{pages}{39670--39679}.
  \DOIprefix\doi{10.1109/ACCESS.2019.2906890}.
\bibitem[{{Sun} et~al.(2019){Sun}, {Zhang}, {Gu}, {Liu}, {Hong}, {Xu}, {Yang},
  and {Gui}}]{superRes1}
\bibinfo{author}{Y.~{Sun}}, \bibinfo{author}{W.~{Zhang}},
  \bibinfo{author}{H.~{Gu}}, \bibinfo{author}{C.~{Liu}},
  \bibinfo{author}{S.~{Hong}}, \bibinfo{author}{W.~{Xu}},
  \bibinfo{author}{J.~{Yang}}, \bibinfo{author}{G.~{Gui}},
\newblock \bibinfo{title}{Convolutional neural network based models for
  improving super-resolution imaging},
\newblock \bibinfo{journal}{IEEE Access} \bibinfo{volume}{7}
  (\bibinfo{year}{2019}) \bibinfo{pages}{43042--43051}.
  \DOIprefix\doi{10.1109/ACCESS.2019.2908501}.
\bibitem[{Chen et~al.(2017)Chen, Li, and Zhu}]{Chen:17}
\bibinfo{author}{S.~Chen}, \bibinfo{author}{C.~Li}, \bibinfo{author}{Y.~Zhu},
\newblock \bibinfo{title}{Sensitivity evaluation of quantitative phase imaging:
  a study of wavelength shifting interferometry},
\newblock \bibinfo{journal}{Opt. Lett.} \bibinfo{volume}{42}
  (\bibinfo{year}{2017}) \bibinfo{pages}{1088--1091}. \URLprefix
  \url{http://ol.osa.org/abstract.cfm?URI=ol-42-6-1088}.
  \DOIprefix\doi{10.1364/OL.42.001088}.
\bibitem[{Feller(1947)}]{cramerOrig}
\bibinfo{author}{W.~Feller},
\newblock \bibinfo{title}{Review: Harald cramer, mathematical methods of
  statistics},
\newblock \bibinfo{journal}{Ann. Math. Statist.} \bibinfo{volume}{18}
  (\bibinfo{year}{1947}) \bibinfo{pages}{136--139}. \URLprefix
  \url{https://doi.org/10.1214/aoms/1177730503}.
  \DOIprefix\doi{10.1214/aoms/1177730503}.
\bibitem[{{Li} and {Zhu}(2017)}]{chengshuaiJSTQE}
\bibinfo{author}{C.~{Li}}, \bibinfo{author}{Y.~{Zhu}},
\newblock \bibinfo{title}{Cramer–rao bound for frequency estimation of
  spectral interference and its shot noise-limited behavior},
\newblock \bibinfo{journal}{IEEE Journal of Selected Topics in Quantum
  Electronics} \bibinfo{volume}{23} (\bibinfo{year}{2017})
  \bibinfo{pages}{410--416}. \DOIprefix\doi{10.1109/JSTQE.2016.2604798}.
\bibitem[{Leon-Garcia(2008)}]{leongarcia08}
\bibinfo{author}{A.~Leon-Garcia}, \bibinfo{title}{Probability, Statistics, and
  Random Processes for Electrical Engineering}, \bibinfo{edition}{third} ed.,
  \bibinfo{publisher}{Pearson/Prentice Hall}, \bibinfo{address}{Upper Saddle
  River, NJ}, \bibinfo{year}{2008}.
\bibitem[{{Li} et~al.(2017){Li}, {Chen}, and {Zhu}}]{chengshuaiMLE}
\bibinfo{author}{C.~{Li}}, \bibinfo{author}{S.~{Chen}},
  \bibinfo{author}{Y.~{Zhu}},
\newblock \bibinfo{title}{Maximum likelihood estimation of optical path length
  in spectral interferometry},
\newblock \bibinfo{journal}{Journal of Lightwave Technology}
  \bibinfo{volume}{35} (\bibinfo{year}{2017}) \bibinfo{pages}{4880--4887}.
  \DOIprefix\doi{10.1109/JLT.2017.2743214}.
\bibitem[{Chen et~al.(2016)Chen, Li, and Zhu}]{Chen2:16}
\bibinfo{author}{S.~Chen}, \bibinfo{author}{C.~Li}, \bibinfo{author}{Y.~Zhu},
\newblock \bibinfo{title}{Low-coherence wavelength shifting interferometry for
  high-speed quantitative phase imaging},
\newblock \bibinfo{journal}{Opt. Lett.} \bibinfo{volume}{41}
  (\bibinfo{year}{2016}) \bibinfo{pages}{3431--3434}. \URLprefix
  \url{http://ol.osa.org/abstract.cfm?URI=ol-41-15-3431}.
  \DOIprefix\doi{10.1364/OL.41.003431}.
\bibitem[{Goodman(2015)}]{goodman_2015}
\bibinfo{author}{J.~W. Goodman}, \bibinfo{title}{Statistical Optics},
  \bibinfo{publisher}{Wiley}, \bibinfo{year}{2015}.
\bibitem[{Chen et~al.(2016)Chen, Ryu, Lee, and Zhu}]{Chen:16}
\bibinfo{author}{S.~Chen}, \bibinfo{author}{J.~Ryu}, \bibinfo{author}{K.~Lee},
  \bibinfo{author}{Y.~Zhu},
\newblock \bibinfo{title}{Swept source digital holographic phase microscopy},
\newblock \bibinfo{journal}{Opt. Lett.} \bibinfo{volume}{41}
  (\bibinfo{year}{2016}) \bibinfo{pages}{665--668}. \URLprefix
  \url{http://ol.osa.org/abstract.cfm?URI=ol-41-4-665}.
  \DOIprefix\doi{10.1364/OL.41.000665}.
\bibitem[{{Carr{\'e}}(1966)}]{carrePaper}
\bibinfo{author}{P.~{Carr{\'e}}},
\newblock \bibinfo{title}{{Installation et utilisation du comparateur
  photo{\'e}lectrique et interf{\'e}rentiel du Bureau International des Poids
  et Mesures}},
\newblock \bibinfo{journal}{Metrologia} \bibinfo{volume}{2}
  (\bibinfo{year}{1966}) \bibinfo{pages}{13--23}.
  \DOIprefix\doi{10.1088/0026-1394/2/1/005}.
\bibitem[{Yariv and Yeh(2009)}]{yariv_yeh_2009}
\bibinfo{author}{A.~Yariv}, \bibinfo{author}{P.~Yeh},
  \bibinfo{title}{Photonics: optical electronics in modern communications},
  \bibinfo{publisher}{Oxford Univ. Press}, \bibinfo{year}{2009}.
\bibitem[{{Gorman} and {Hero}(1990)}]{cCRB}
\bibinfo{author}{J.~D. {Gorman}}, \bibinfo{author}{A.~O. {Hero}},
\newblock \bibinfo{title}{Lower bounds for parametric estimation with
  constraints},
\newblock \bibinfo{journal}{IEEE Transactions on Information Theory}
  \bibinfo{volume}{36} (\bibinfo{year}{1990}) \bibinfo{pages}{1285--1301}.
  \DOIprefix\doi{10.1109/18.59929}.
\bibitem[{Chollet et~al.(2015)}]{chollet2015keras}
\bibinfo{author}{F.~Chollet}, et~al., \bibinfo{title}{Keras},
  \bibinfo{howpublished}{\url{https://keras.io}}, \bibinfo{year}{2015}.
\bibitem[{Kingma and Ba(2015)}]{Kingma2015AdamAM}
\bibinfo{author}{D.~P. Kingma}, \bibinfo{author}{J.~Ba},
\newblock \bibinfo{title}{Adam: A method for stochastic optimization},
\newblock \bibinfo{journal}{CoRR} \bibinfo{volume}{abs/1412.6980}
  (\bibinfo{year}{2015}).

\end{thebibliography}

\end{document}